\documentclass[aps,preprint]{revtex4}%
\usepackage{amsfonts}
\usepackage{amsmath}
\usepackage{amssymb}
\usepackage{graphicx}
\usepackage{epstopdf}%
\begin{document}
\preprint{ }
\title{Radiation-assisted magnetotransport in
two-dimensional electron gas systems:
appearance of zero resistance states}
\author{Abdullah Yar*}
\email{abdullahyardawar@gmail.com}
\author{Kashif Sabeeh\dag}
\affiliation{*Department of Physics, Kohat University of Science and Technology,
Kohat-26000, Khyber Pakhtunkhwa, Pakistan}

\affiliation{\dag Department of Physics, Quaid-i-Azam University, Islamabad 45320 Pakistan}

\begin{abstract}
Zero-Resistance States (ZRS) are normally associated with superconducting and quantum Hall
phases. Experimental detection of ZRS in two-dimensional electron gas (2DEG) systems irridiated
by microwave(MW) radiation in a magnetic field has been quite a surprise. We develop a
semiclassical transport formalism to explain the phenomena. We find a sequence of Zero-Resistance
States (ZRS) inherited from the suppression of Shubnikov-de Haas (SdH) oscillations under the
influence of high-frequency and large amplitude microwave radiation. Furthermore, the ZRS are well
pronounced and persist up to broad intervals of magnetic field as observed in experiments on
microwave illuminated 2DEG systems.

\end{abstract}
\startpage{01}
\endpage{02}
\maketitle

\section{Introduction}

Magnetotransport provides important information on Fermi surface
characteristics, disorder and localization mechanisms in low-dimensional
electron systems. A comprehensive review of the electronic properties and
electronic transport in two-dimensional electron gas (2DEG) systems was
presented by Ando, Fowler and Stern~\cite{AndoRMP54.437} in early eighties.
The discovery of Quantum Hall effects around the same time led to the realization
that, in strong magnetic field, quantization effects lead to novel transport features~\cite{KlitzingRMP58.519,LaughlinRMP71.863,StormerRMP71.875,TsuiRMP71.891}.
In the late eighties, it was found that in moderately strong magnetic fields; when
quantization effects are absent but cyclotron dynamics is present; interesting set
of phenomena occur. In this regard, it was observed that commensurability oscillations
in the magnetoresistance occur in periodically modulated 2DEG
systems~\cite{VasilopoulosPRL2120,PeetersPRB46-4667,
ZhangPRB41.12850}. In particular, periodic oscillations in
1/B (B is the applied magnetic field) is observed in the magnetoresistance of
a two-dimensional electron gas (2DEG) subjected to
weak~\cite{WeissEPL8-179,WeissPRL66.2790,WinklerPRL62-1177,
AlvesJPCM1-8257,Beenakkerprl2020,PfannkuchePRB46.12606} and
strong~\cite{Beton42-9229,GvozdikovPRB75.115106,ShiPRB53.12990} periodic potential.
In addition to magnetoresistance, magnetoplamons in these systems have also been
investigated,~\cite{KushwahaSSR41.1-416}
and references therein. In a series of experiments carried out in 2001-2003,
it was found that a 2DEG subjected to microwave radiation in an applied magnetic field
yields even richer physics. When a high mobility 2DEG is irradiated by microwave radiation
in a weak magnetic field, the longitudinal magnetoresistance exhibits giant oscillations.
This was the discovery of Microwave Induced Resistance Oscillations (MIRO)~\cite{StudenikinPRB165321,HaltePRB121301,DuIJMP81.3465,Dorozhkinprb201306,
MikhailovPRB70.165311,StudenikinSol.St.Comm.129,KukushkinPRB121306,YeAPL79.2193,
ZudovPRB64.201311,LeiPRL91.226805,LeiPRB72.075345,BykovJETPL84.391}.
A significant feature of these studies has been the
observation of Zero-Resistance States (ZRS) in these systems; the lower order
minima in MIRO go all the way to zero~\cite{Mani
Nat.420-646,ManiAPL-85-4962,Zudovprl046807,Zudovprb041304,
Boganprb235305,Manisrep03478,WiedmannPRL105.026804}. Observation of ZRS was quite a
surprise; eventhough longitudinal resistance exhibits ZRS in integer quantum
Hall systems but the magnetic field required here is smaller by a factor of
50. Unlike Quantum Hall phenomena, the vanishing of longitudinal magnetoresistance
does not lead to quantization of Hall resistance. This led to
the understanding that weak Landau quantization and weak microwave radiation
can significantly alter the transport properties of a 2DEG. This discovery
opened the field of nonequilibrium transport in high Landau levels~\cite{DmitrievRMP84.1709}.
Several explanations have been put
forward~\cite{Robinsonprl036804,Kennett195420,VavilovPRB70.161306, InarreaPRL94.016806,
WiedmannPRL105.026804,BykovJETPL84.391}. Most of the theoretical work relies
on the combined effect of Landau quantization and applied fields on momentum
relaxation due to impurity scattering with in a Landau band; alternatively
experimental results are explained on the basis of redistribution of electrons
in a disorder broadened Landau band due to interaction with microwaves.

The mechanism responsible for the appearance of ZRS in a 2DEG in the presence
of both MW radiation and an external magnetic field is still far from settled.
In this work, we will investigate whether it is possible to find an
explanation of this phenomenon with in a single particle semiclassical
picture. In this regard, we will focus on the effect of plane polarized
electromagnetic radiation on the cyclotron motion of electrons in a 2DEG
system. Cyclotron motion of electrons in a 2DEG has been extensively studied
in~\cite{JohnPRB125303,WinklerPRB205314,JohnPRB085323} . We base our study
on~[\onlinecite{Beenakkerprl2020,Kennett195420}] and extend it to include the
effects of electromagnetic radiation. In particular, we find that cyclotron
trajectories are significantly modified under the influence of radiation.
Further, commensurability oscillations in the magnetoresistivity of 2DEG are
induced by radiation. Interestingly, we find a sequence of zero-resistance
states (ZRS) inherited from the suppression of Shubnikov-de Haas oscillations
(SdHO) by high-frequency and large amplitude microwave radiation. The ZRS are
well pronounced and persist up to broad intervals of magnetic field as
observed in many experiments on microwave illuminated 2DEG~\cite{Mani
Nat.420-646,DuIJMP81.3465}. Moreover, the formation of ZRS strongly depends on
the frequency of MW radiation and disappear at low frequencies. This fact is
further confirmed by investigating a range of MW frequency where the system is
completely driven to ZRS.\newline\indent The paper is organized as follows: In
Sec.~\ref{Sec:Hamiltonian}, the model Hamiltonian of our system which is a
2DEG in the presence of a perpendicular magnetic field and microwave radiation
is introduced. The eigenstates of the system in the absence of radiation are
determined. The investigation of cyclotron motion of electrons in the presence
of both MW radiation and an external magnetic field is formulated in the
framework of Heisenberg equation of motion technique and the semiclassical
formalism is derived from the full quantum description. Moreover, the
influence of microwave radiation on the magnetic field-assisted dynamics is
studied in detail.\newline\indent Sec.~\ref{Sec:Formalism} is dedicated to the
formulation of magnetoresistivity by finding the enhancement in diffusion
coefficients using the drift velocity of electrons.\newline\indent In
Sec.~\ref{Sec:Results} the results based on our model are discussed. The
different limiting cases are analyzed in detail. Finally, conclusions are
drawn in Sec.~\ref{Sec:Conc}.

\section{Model Hamiltonian}

\label{Sec:Hamiltonian}

The single particle Hamiltonian of an electron in a 2DEG system (in the $xy$
plane) in the presence of electromagnetic (MW) radiation polarized along the
$x-$ direction in the plane subjected to a perpendicular magnetic field is
given by
\begin{equation}
\hat{\mathcal{H}}=\frac{\hat{\pi}^{2}}{2m^{\ast}}+e\hat{x}E_{0}\cos(\omega t),
\label{eq:Hamiltonian}%
\end{equation}
where $\mathbf{{\hat{\pi}}}=\left(  \pi_{x},\pi_{y}\right)  =\left(  \hat
{p}+e\mathbf{A}\right)  $ is the two-component kinetic momentum with the
canonical momentum operator $\hat{p}$, and $\mathbf{A}$ is the vector
potential given by $\mathbf{A}=(0,xB,0)$ in the Landau gauge. Moreover,
$m^{\ast}$ is the effective mass of electron in 2DEG. The second term in the
above Hamiltonian represents interaction of the electromagnetic wave (MW) with
the electron. The constant $E_{0}$ is the amplitude of the electric field of
the electromagnetic wave and $\omega$ is its angular frequency. In the above
Hamiltonian, we have neglected the spatial variation in the electric field of
the wave~\cite{GriffithsQM}. This approximation is reasonable if we consider
the MW with wavelength larger than the diameter of the cyclotron orbit
$d_{c}=2r_{c}=2k_{F}l^{2}$, with $k_{F}$ being the Fermi momentum and
$l=\sqrt{\hbar/eB}$ the magnetic length.

In the absence of MW, the normalized eigenstates of the system are given by
\begin{equation}
\psi_{nk_{y}}(x,y)=\frac{1}{\sqrt{L_{y}}}e^{-ik_{y}y}\varphi_{n}(x),
\label{eq:Basis}%
\end{equation}
where $L_{y}$ is the length of the sample in the $y$-dimension. The functions
$\varphi_{n}(x)$ represents the eigenstates of harmonic oscillator with
guiding centre at $x_{0}$ described by
\begin{equation}
\varphi_{n}(x)=\frac{1}{\sqrt{2^{n}n!\sqrt{\pi}l}}e^{-\frac{1}{2}\left(
\frac{x-x_{0}}{l}\right)  ^{2}}H_{n}\left(  \frac{x-x_{0}}{l}\right)  ,
\end{equation}
where $H_{n}(x)$ is the $n$th-order Hermite polynomial, $x_{0}=l^{2}k_{y}$ is
the centre of cyclotron orbit. In the above expression $n=0,\pm1,\pm2,...$
characterizes the Landau levels and $k_{y}$ is the electron wave number with
the translational invariance in the $y$ direction. The quantum number $k_{y}$
is conveniently determined by periodic boundary condition as
\begin{equation}
k_{y}=\frac{2\pi}{L_{y}}n. \label{Eq:Bound:Cond}%
\end{equation}
The maximum value of $n$ can be specified by the condition that the centre of
the cyclotron orbit should be within the sample: $0<x_{0}<L_{x}$, where
$L_{x}$ is the dimension of the sample in the $x$-dimension. Alternatively
\begin{equation}
|k_{y}|<\frac{L_{x}}{l^{2}}=\frac{|eB|}{\hbar}L_{x}. \label{Eq:Bound:Cond1}%
\end{equation}

\subsection{Quantum description of cyclotron motion in a 2DEG}

\label{Subsec:Scattering}

In this section, we analyze the cyclotron motion of a charged particle in the
2DEG illuminated by MW within a quantum mechanical approach. The time
evolution of the cyclotron trajectories is determined using Heisenberg
equation of motion. As a consequence, the time evolution of the position
operator $\hat{r}$ in Heisenberg picture reads
\begin{align}
\frac{d\hat{r}}{dt}=\frac{i}{\hbar}\left[  \hat{\mathcal{H}},\hat{r}\right]  ,
\end{align}

After straightforward calculations the above equation of motion yields
\begin{align}
\label{Eq:position}\hat{r}(t)  &  =\hat{r}(0)+\frac{\hat{\pi}(0)}{m^{\ast}%
}\frac{1-e^{-i\left(  \omega_{c}t+\varphi_{0}\right)  }}{i\omega_{c}}
+\frac{eE_{0}}{m^{\ast} \left(  \omega_{c}^{2}-\omega^{2}\right)
}e^{-i\left(  \omega_{c}t+\varphi_{0}\right)  }\nonumber\\
&  -\frac{eE_{0}}{m^{\ast}\left(  \omega_{c}^{2}-\omega^{2}\right)  }\left[
\cos\left(  \omega t+\frac{\omega}{\omega_{c}}\varphi_{0}\right)
-i\frac{\omega_{c}}{\omega}\sin\left(  \omega t+\frac{\omega}{\omega_{c}%
}\varphi_{0}\right)  \right]  ,
\end{align}

where $\omega_{c}=\frac{|eB|}{m^{\ast}}$ is the cyclotron frequency and
$\vec{r}(0)$ specify the initial coordinates of the centre of the cyclotron
orbit which commute with the Hamiltonian of the system and consequently
remains constant in time. The constant phase $\varphi_{0}$ locates the initial
position of the particle in the cyclotron orbit. Eq.~\eqref{Eq:position}
reveals that the cyclotron trajectories are significantly affected by the
microwave (MW). In order to analyze the MW-assisted dynamics of the particle
in a magnetic field we need to evaluate the expectation values of the time
dependent position operators. Using the complex notations $\hat{r}=\hat
{x}-i\hat{y}$ and $\hat{\pi}=\hat{\pi}_{x}-i\hat{\pi}_{y}$, the $x$-component
of the cyclotron motion can be expressed as
\begin{align}
\label{Eq:positionx}\hat{x}(t)  &  =\hat{x}(0)+\frac{\hat{\pi}_{x}(0)}%
{m^{\ast}\omega_{c}}\sin(\omega_{c}t+\varphi_{0}) +\frac{\hat{\pi}_{y}%
(0)}{m^{\ast}\omega_{c}} \left[  \cos(\omega_{c}t+\varphi_{0})-1\right]
\nonumber\\
&  +\frac{eE_{0}}{m^{\ast}\left(  \omega_{c}^{2}-\omega^{2}\right)  }%
\cos(\omega_{c}t+\varphi_{0}) -\frac{eE_{0}}{m^{\ast}\left(  \omega_{c}%
^{2}-\omega^{2}\right)  }\cos\left(  \omega t+\frac{\omega}{\omega_{c}}%
\varphi_{0}\right)  .
\end{align}
Similarly, the $y$-component of the cyclotron motion can be described in the
form
\begin{align}
\label{Eq:positiony}\hat{y}(t)  &  =\hat{y}(0)-\frac{\hat{\pi}_{x}(0)}%
{m^{\ast}\omega_{c}}\left[  \cos(\omega_{c}t+\varphi_{0})-1\right]
+\frac{\hat{\pi}_{y}(0)}{m^{\ast} \omega_{c}}\sin(\omega_{c}t+\varphi
_{0})\nonumber\\
&  +\frac{eE_{0}}{m^{\ast}\left(  \omega_{c}^{2} -\omega^{2}\right)  }%
\sin(\omega_{c}t+\varphi_{0}) -\frac{eE_{0}\omega_{c}}{m^{\ast}\omega\left(
\omega_{c}^{2}-\omega^{2}\right)  }\sin\left(  \omega t+\frac{\omega}%
{\omega_{c}}\varphi_{0}\right)  .
\end{align}

\subsection{Semiclassical formulation of cyclotron motion}

Eqs.~\eqref{Eq:positionx} and~\eqref{Eq:positiony} give the full quantum
mechanical description of the cyclotron motion in two dimensional electron
systems in the presence of an external perpendicular magnetic field when the
system is irridiated by MW. However, in order to obtain analytic results we
develop a semiclassical formalism. In this regard, we are interested in the
classical expectation values of the cyclotron trajectories which are obtained
by replacing the operators by their corresponding classical variables. The
analytic expressions for the electron dynamics in external magnetic field and
MW radiation can be obtained in the semiclassical regime specified by the
criterion, $k_{\text{F}}l\gg1$, where $k_{\text{F}}$ is the Fermi wave vector
given by $k_{\text{F}}=\sqrt{2\pi n_{e}}$ with $n_{e}$ being the electron
density. Semiclassical results can be obtained from quantum mechanical
results by ignoring the quantum fluctuations of the operators corresponding
to dynamical variables. In our case, one can derive the semiclassical results from the full
quantum mechanical equations~\eqref{Eq:positionx} \&~\eqref{Eq:positiony} by
treating the operators $\hat{x}(t)$, $\hat{x}(0)$, $\hat{y}(t)$, $\hat{y}(0)$,
$\hat{\pi}_x$, and $\hat{\pi}_y$ as classical variables. Consequently,
in the semiclassical limit expectation value of the
$x$-component of the position operator can be expressed as
\begin{align}
x(t)  &  =x(0)+\kappa_{x}l^{2}\sin{(\omega_{c}t+\varphi_{0})}+2\kappa_{y}%
l^{2}\left[  \cos(\omega_{c}t+\varphi_{0})-1\right]
\nonumber\label{Eq:Cyclotronx}\\
&  +\frac{eE_{0}}{m^{\ast}\left(  \omega_{c}^{2}-\omega^{2}\right)  }%
\cos(\omega_{c}t+\varphi_{0})-\frac{eE_{0}}{m^{\ast}\left(  \omega_{c}%
^{2}-\omega^{2}\right)  }\cos\left(  \omega t+\frac{\omega}{\omega_{c}}%
\varphi_{0}\right)  ,
\end{align}
In a similar way, the expectation value of the $y$-component of the position
operator can be written as
\begin{align}
y(t)  &  =y(0)-l^{2}k_{x}\left[  \cos(\omega_{c}t+\varphi_{0})-1\right]
+2l^{2}k_{y}\sin(\omega_{c}t+\varphi_{0})\nonumber\label{Eq:Cyclotrony}\\
&  +\frac{eE_{0}}{m^{\ast}\left(  \omega_{c}^{2}-\omega^{2}\right)  }%
\sin(\omega_{c}t+\varphi_{0})-\frac{eE_{0}\omega_{c}}{m^{\ast}\omega\left(
\omega_{c}^{2}-\omega^{2}\right)  }\sin\left(  \omega t+\frac{\omega}%
{\omega_{c}}\varphi_{0}\right)  ,
\end{align}
where $k_{y}$ is the $y$-component of the electron wave vector given by
Eq.~\eqref{Eq:Bound:Cond} and $k_{x}$ is its $x$-component which is determined
by the relation, $k_{x}=\sqrt{k_{\text{F}}^{2}-k_{y}^{2}}$.

\begin{figure}[ptb]
\begin{center}
\includegraphics[%width=\columnwidth
height=4.8in,
width=6.2in
]{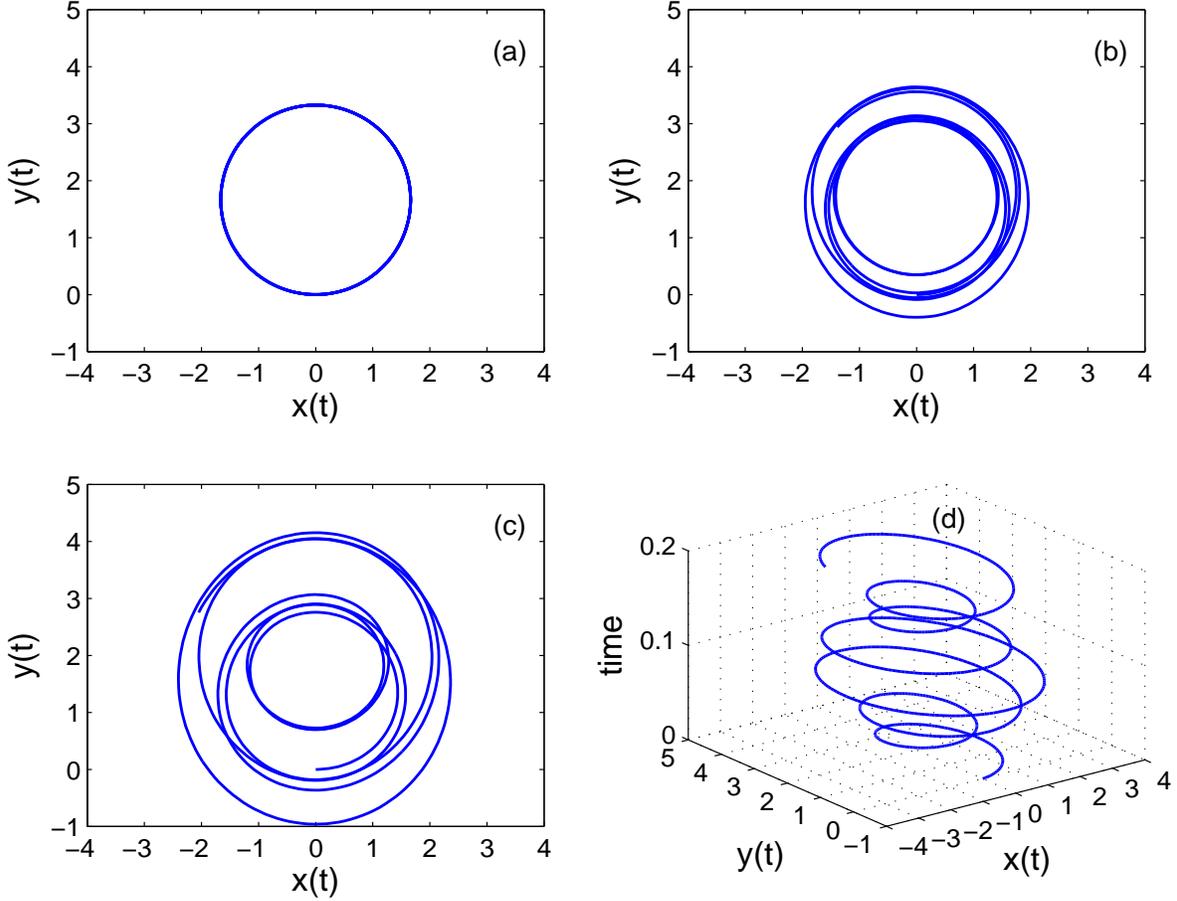}
\end{center}
\caption{Semiclassical cyclotron orbit dynamics of the particle, (a) without
microwave radiations (MWs), whereas (b), (c) and (d) with MWs. The $x$- and
$y$- coordinates are measured in $\mu\text{m}$. The experimentally relevant
set of parameters used are: the frequency of MW is $f=25\ \text{GHz}$, charge
carrier density is $n_{e}= 65\times10^{14}\ \text{m}^{-2}$ and the effective
mass of the electron is $m^{\ast}=0.068\ m_{0}$. Length of the system is
$L_{x}=6\ m\text{m}$ and its width is $L_{y}=6\ m\text{m}$. The amplitude of
MW electric field is $E_{0}=2\times10^{3}\ \text{V}\text{m}^{-1}$ for (b) and
$E_{0}=4.5\times10^{3}\ \text{V}\text{m}^{-1}$ for (c) and (d). The external
magnetic field is $B=0.08\ \text{T}$. The initial coordinates and phase are
$x(0)=0$, $y(0)=0$, and $\varphi_{0}=0$, respectively. In (d) we have
demonstrated the time evolution of the cyclotron orbit where the time is
measured in units of nano second. The simulation time is always $t=10\ \pi
/\omega$.}%
\label{fig1}%
\end{figure}

\begin{figure}[ptb]
\begin{center}
\includegraphics[%width=\columnwidth
height=4.158in,
width=5.3722in
]{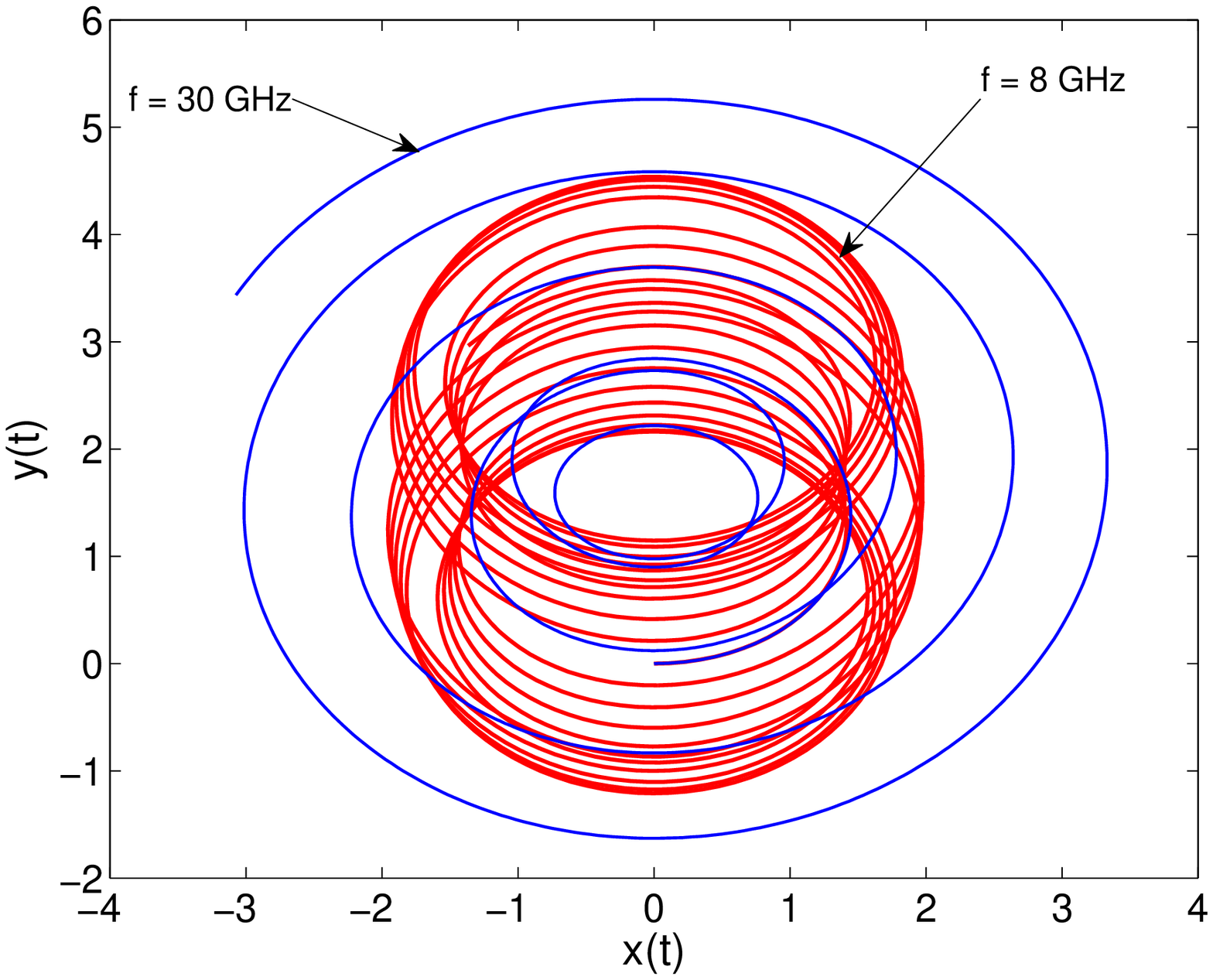}
\end{center}
\caption{Cyclotron orbit dynamics of the electron under the influence of
microwave radiations for two different frequencies. The amplitude of SW
electric field is $E_{0}=4.5\times10^{3}\ \text{V}\text{m}^{-1}$ and the other
parameters are the same as used in Fig.~\ref{fig1}.}%
\label{fig2ab}%
\end{figure}

In Fig.~\ref{fig1} we have shown the dynamics of the electronic classical
orbit evaluated in terms of the expectation values of the cyclotron
trajectories in the basis described by Eq.~\eqref{eq:Basis}.
It is evident from this figure that the cyclotron trajectories are strongly
modified by the microwave radiation. This modification in the cyclotron orbit
depends on the amplitude of MW electric field which specifies the coupling of
the radiation to the electronic degrees of freedom.

In order to independently investigate the effect of MW frequency on cyclotron
motion, we plot the results in Fig.~\ref{fig2ab}.
The comparison of thick red curve and thin blue curve shows that larger shift
in guiding center takes place at higher frequency. In summary, a shift is
produced in the guiding center of the electronic cyclotron orbit under the
effect of MW radiation which in turn affects the transport properties of the system.

\section{Magnetoresistivity}

\label{Sec:Formalism}

We adopt the semiclassical approach for evaluating magnetoresistivity
developed by Beenakker~\cite{Beenakkerprl2020} and Kennet~\cite{Kennett195420}%
. In order to simplify the analysis, we take the electric dipole moment
$\vec{\mu}=e\vec{x}$ of the electron and the electric field $\vec{E}_{0}$ of
the microwave radiation to be parallel polarized. As a result, Lorentz force
is experienced by the electron that causes drift ($\vec{E}_{0}\times\vec{B}$)
of the guiding center of the cyclotron orbit in the transverse direction. The
drift velocity of the electron guiding center in the transverse direction can
be described to the lowest order of the MW radiation field as
\begin{equation}
v_{y}(t)\approx\frac{E_{0}}{B}\cos[qx(t)-\omega t],\quad v_{x}(t)=0,
\label{Eq:drift velocity}%
\end{equation}
where $x(t)$ is the instantaneous position of the electron in cyclotron motion
and $q$ is the wave number of the MW. Due to the transverse velocity, the
diffusion coefficient tensor $D_{yy}$ in the transverse direction is enhanced.
This diffusion coefficient can be determined from the autocorrelation function
of the electron velocities by taking average over all the particle
trajectories and scattering events
\begin{equation}
D_{yy}=\int_{0}^{\infty}\int_{0}^{2\pi}\int_{0}^{2\pi}dte^{-t/\tau}%
\frac{d\varphi_{0}}{2\pi}\frac{d\xi}{2\pi}v_{y}(t)v_{y}(0), \label{Eq:DCoef}%
\end{equation}
where $\xi=qx(0)$. Once the transverse diffusion tensor $D_{yy}$ is known, one
can find the longitudinal resistivity tensor $\rho_{xx}$ using Einstein
diffusion relation~\cite{Beenakkerprl2020}
\begin{equation}
\frac{\rho_{xx}}{\rho_{0}}=\frac{D_{yy}}{D_{0}}, \label{Eq:Resistivity}%
\end{equation}
where $\rho_{0}$ is the Drude resistivity in zero magnetic field and $D_{0}$
represents the unperturbed diffusion coefficient. For a 2DEG the diffusion
coefficient in the presence of magnetic field is given by $D_{0}=r_{c}%
^{2}/2\tau$ with $r_{c}=v_{\text{F}}/\omega_{c}$ being the classical cyclotron
radius and $\tau$ the transport relaxation time. Using
Eqs.~\eqref{Eq:Cyclotronx},~\eqref{Eq:drift velocity},~\eqref{Eq:DCoef},~\eqref{Eq:Resistivity}
and the Bessel function identities~\cite{Bessel} one can find
\begin{widetext}
\begin{equation}
\begin{split}
\frac{\rho_{xx}}{\rho_0}=\left(\frac{\tau e E_0 v_\text{F}}{2\epsilon_\text{F}} \right)^2\sum^\infty_{n=-\infty}\sum^\infty_{m=-\infty}
\sum^\infty_{k=-\infty}\sum^\infty_{s=-\infty}
\frac{J^2_n\left(qk_xl^2\right)J^2_m\left(2qk_yl^2\right)J^2_k\left[\frac{eE_0q}{m^\ast\left(\omega^2_c
-\omega^2\right)}\right]J^2_s\left[\frac{eE_0q}{m^\ast\left(\omega^2_c
-\omega^2\right)}\right]}{1+\left[\omega(1+s)-\omega_c\left(n+m+k\right)\right]^2\tau^2},
\end{split}
\label{Eq:Mag Resistivity}
\end{equation}
\end{widetext}
where $v_{\text{F}}$ is the Fermi velocity, $\epsilon_{\text{F}%
}$ is the Fermi energy of the electron and $J_{n}(x)$ is the $n$th-order
Bessel function of the first kind.

\section{Results and discussions}

\label{Sec:Results}

\begin{figure}[th]
\begin{center}
\includegraphics[%width=\columnwidth
height=4.158in,
width=5.3722in
]{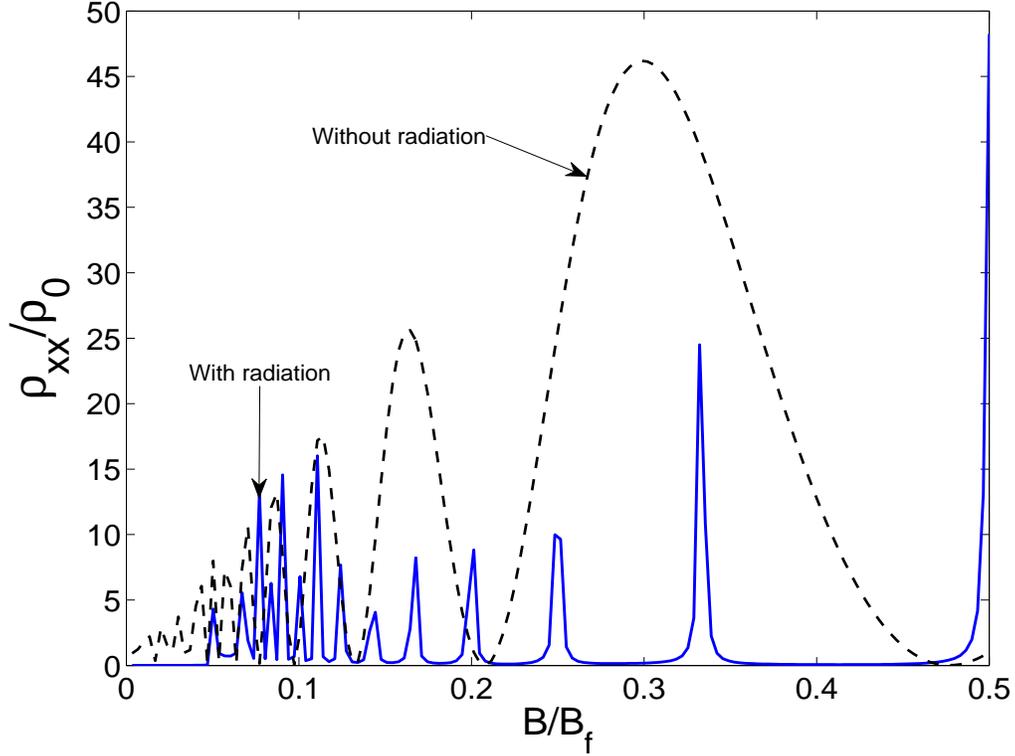}
\end{center}
\caption{Magnetoresistivity of GaAs based two dimensional electron gas (2DEG).
The dashed black curve represents magnetoresistivity in static limit
($\omega\rightarrow0$), whereas the blue curve denotes the dynamical
magnetoresistivity in the presence of microwave radiation. The wavelength of
MW radiation is $\lambda=4\ \mu\text{m}$, the amplitude of MW electric field
is $E_{0}=4.3\times10^{3}\ \text{V}\text{m}^{-1}$, whereas its frequency is
$f=75\ \text{GHz}$. The other parameters are the same as used in
Fig.~\ref{fig1}.}%
\label{fig3}%
\end{figure}

In this section, we illustrate the results of our model and discuss the
various features arising in magnetotransport, in particular, the formation of
zero-resistance states. We use the parameters set that is in the
experimentally relevant range given in
Refs.~[\onlinecite{Mani Nat.420-646,ManiAPL-85-4962,HaltePRB121301}]. In
Fig.~\ref{fig3}, we demonstrate the resistivity ratio given by
Eq.~\eqref{Eq:Mag Resistivity} as a function of $B/B_{f}$, where $B_{f}=2\pi
fm^{\ast}/e$.

A detailed analysis of this plot reveals that in the static limit
($\omega\rightarrow0$) the magnetoresistivity exhibits the usual Shubnikov-de
Haas oscillations (SdHO), see the black dashed curve in Fig.~\ref{fig3}. This
effect arises from Landau level quantization in the magnetic field. This
feature of the system can be described by writing
Eq.~\eqref{Eq:Mag Resistivity} in the form

\begin{equation}
\frac{\rho _{xx}}{\rho _{0}}\approx \left(\frac{\tau e E_0 v_\text{F}}
{2\epsilon_\text{F}} \right)^2
\sum_{n,m=-\infty }^{\infty }\frac{J_{n}^{2}\left(
qk_{x}l^{2}\right) J_{m}^{2}\left( 2qk_{y}l^{2}\right) }{1+\left( n+m\right)
^{2}\omega _{c}^{2}\tau ^{2}},
\end{equation}%
which in the asymptotic limit closely resembles the Weiss Oscillations in
the diffusion contribution to the resistivity
~\cite{WinklerPRL62-1177,GerhardtsPRB5192,VasilopoulosPRL2120,PeetersPRB46-4667}.
The scenario changes and interesting features appear when the system is
illuminated by high-frequency microwave radiation. The Shubnikov-de Haas
oscillations (SdHO) are suppressed and even the resistivity of the system
vanishes within certain intervals of the magnetic field. The states
responsible for zero resistivity are known as zero-resistance states (ZRS).
The mechanism of suppressed resistivity and the consequent zero-resistance
states can be better understood by the following analytic analysis of the
above equation: the suppression of SdHO is enhanced by the interference
effects between periodically oscillating functions. In order to understand
this mechanism we consider the asymptotic behavior of the Bessel function in the limits,
$qk_{x}l^{2},\ qk_{y}l^{2},\ \frac{eE_{0}q}{%
m^{\ast }\left( \omega _{c}^{2}-\omega ^{2}\right) }\gg 1$. Under the above
approximation Eq.~\eqref{Eq:Mag Resistivity} can be recast into the form ($m\neq 0$)

\begin{widetext}
\begin{equation}
\begin{split}
\frac{\rho_{xx}}{\rho_0}&\approx \frac{16\tau^2 \left(\omega^2_c
-\omega^2\right)^2}{\pi^4q^4l^4v^2_\text{F}}\sum_{n,m}\sum_{k,s}
\frac{\cos^2\left(qk_xl^2-\frac{n\pi}{2}-\frac{\pi}{4}\right)
\cos^2\left(2qk_yl^2-\frac{m\pi}{2}-\frac{\pi}{4}\right)}
{k_xk_y\left\{1+\left[\omega(1+s)-\omega_c\left(n+m+k\right)\right]^2
\tau^2\right\}}\nonumber\\&\times\cos^2\left[\frac{eE_0q}{m^\ast
\left(\omega^2_c-\omega^2\right)}-\frac{k\pi}{2}-\frac{\pi}{4}\right]
\cos^2\left[\frac{eE_0q}{m^\ast\left(\omega^2_c-\omega^2\right)}
-\frac{s\pi}{2}-\frac{\pi}{4}\right],
\end{split}
\label{Eq:Interference}
\end{equation}
\end{widetext}

Minima in the resistivity can be obtained if at least one of
the following conditions is fulfilled:
\begin{align}
& qk_{x}l^{2}\approx \frac{\pi }{2}\left( n+\frac{3}{2}\right) ,\frac{\pi }{2}
\left( n-\frac{1}{2}\right)  \text{or}\ 2qk_{y}l^{2}\approx \frac{\pi }{2}\left( m-
\frac{1}{2}\right) ,\frac{\pi }{2}\left( m+\frac{3}{2}\right) \nonumber\\&
\text{or} \frac{eE_{0}q}{m^{\ast
}\left( \omega _{c}^{2}-\omega ^{2}\right) }\approx \frac{\pi }{2}\left( k+
\frac{3}{2}\right) ,\frac{\pi }{2}\left( k-\frac{1}{2}\right) ,\newline
\frac{\pi }{2}\left( s-\frac{1}{2}\right) ,\frac{\pi }{2}\left( s+\frac{3}{2}\right).
\end{align}
Due to the alternating behavior of the integers $n,m,s,$ and $k$,
the above conditions can often be satisfied. That is why the zero-resistance
states are very pronounced and persist up to broader intervals of the
magnetic field compared to the one pointed out in Refs.~[~%
\onlinecite{InarreaPRL94.016806,Robinsonprl036804,Kennett195420}]. Furthermore, the most
pronounced ZRS (right side of Fig.~\ref{fig3}) occurs at about $4/9\ B_f$ which is in good
agreement with the occurrence of ZRS observed in Ref.~[\onlinecite{Mani Nat.420-646}].
The next ZRS in our model is near $4/13\ B_f$ and so on.
In summary, the theoretical
model presented in this paper predicts the dynamics of a particle to be
composed of the product of many harmonics. When all these harmonics are in
phase, the drift is enhanced. However, if at least any two harmonics become
out of phase, they cancel the effects of each other and the resistivity
takes a minimum value. Moreover, in the situation investigated here, the
domain of oscillations for in phase/out of phase harmonics is broader due to
the alternating behavior of the Bessel functions. Based on the result of
magnetoresistivity given by Eq.~\eqref{Eq:Mag Resistivity}, our model
predicts different regimes: (i) For weak enough coupling of electromagnetic
wave to electron in the limit $\left(\omega^2_c-\omega^2\right)\gg \frac{eE_0q}{m^\ast}$,
one can approximate $J_{k}(x)\approx \frac{x^{k}}{2^{k}k!}$ in Eq.~%
\eqref{Eq:Mag Resistivity}. Hence, the contribution of microwave radiation does not oscillate
and the resistivity of the system exhibits usual SdH oscillations which stems
qualitatively from the effects of external magnetic field alone.
\begin{figure}[ptb]
\begin{center}
\includegraphics[
height=4.158in,width=5.3722in]
{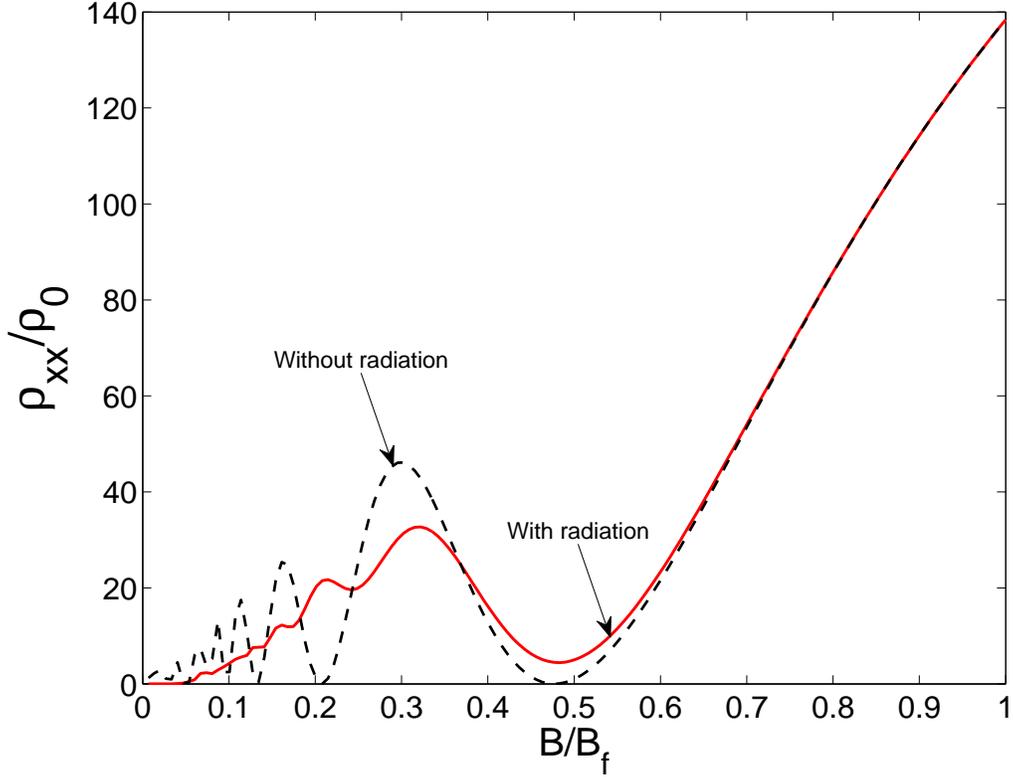}%
\caption{Magnetoresistivity of 2DEG for two different cases. The black dashed curve
represents the resistivity in the static limit, whereas the red curve shows
the dynamical resistivity for microwave radiation
frequency $f=0.1\ \text{GHz}$. In the limit of large magnetic field there is
a good agreement between the static and dynamical resistivities. The
other parameters are the same as used in Fig.~\protect\ref{fig3}.}\label{fig4}
\end{center}
\end{figure}
(ii) In the limit of strong magnetic field, we can again make the above approximation for
Bessel function under the conditions $\left(\omega^2_c-\omega^2\right)\gg \frac{eE_0q}{m^\ast}$
and $qk_xl^2,\ 2qk_yl^2\ll 1$, the commensurability oscillations of the resistivity
are significantly suppressed. That is why the
zero-resistance states are very pronounced at large magnetic field.
(iii) At low magnetic field the commensurability oscillations in the
magnetoresistivity of the system are suppressed because the oscillating
Bessel functions average out to a constant.
(iv) At low frequency of the MW radiation, the system exhibits
SdH oscillations in the resistivity, see Fig.~\ref{fig4}. The same results are approached in
the high magnetic field limit discussed above.
\begin{figure}[ptb]
\begin{center}
\includegraphics[height=4.158in,width=5.3722in]
{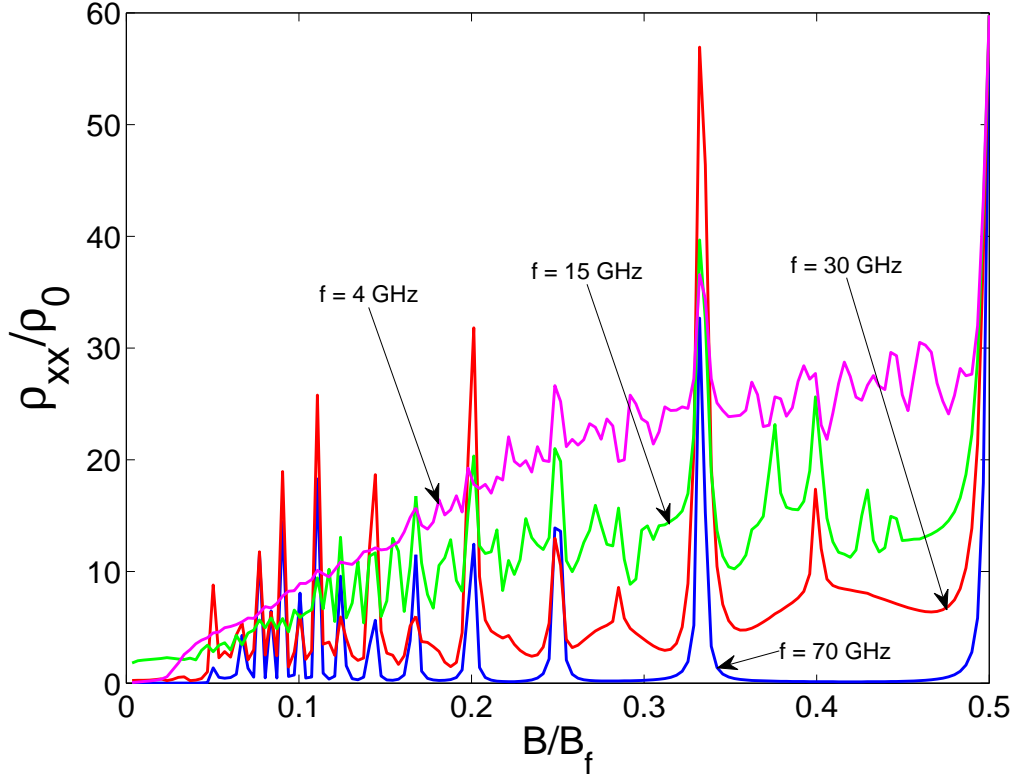}%
\caption{Magnetoresistivity of two dimensional electron gas
for different values of microwave radiation frequencies. The other
parameters are the same as used in Fig.~\protect\ref{fig3}.}\label{fig5}
\end{center}
\end{figure}

\begin{figure}[ptb]
\begin{center}
\includegraphics[height=4.158in,width=5.3722in]
{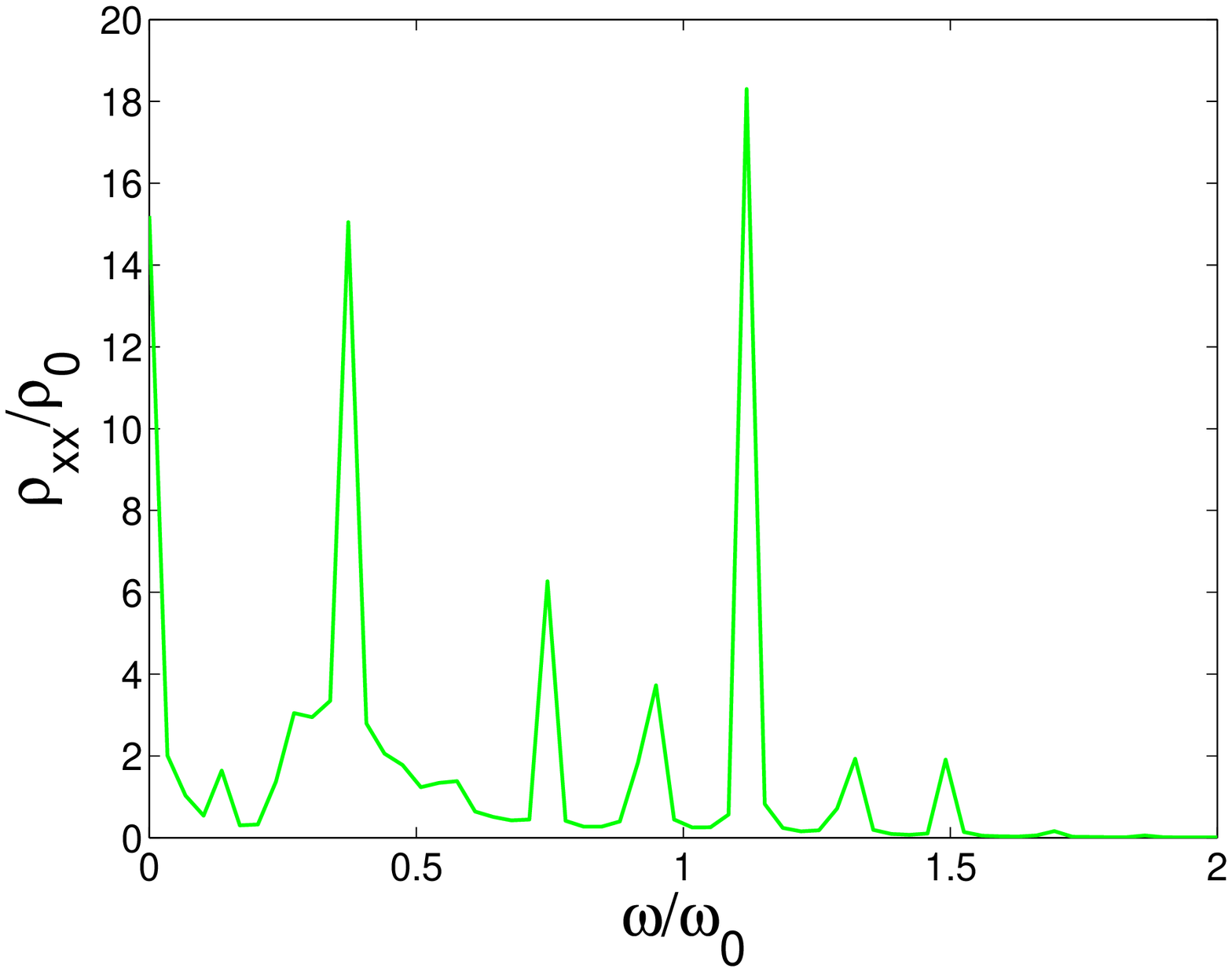}%
\caption{Magnetoresistivity of 2DEG for constant microwave radiation
frequency $f_{0}=66\ \text{GHz}$ and magnetic field $B=0.03\ \text{T}$. The
other parameters are the same as used in Fig.~\protect\ref{fig3}.}\label{fig6}
\end{center}
\end{figure}
Moreover, the expression given by Eq.~\eqref{Eq:Mag Resistivity} reveals that the system
shows a sequence of resonances at those values of the magnetic field which
can fulfill the condition, $\omega (1+s)\approx (n+m+k)\omega _{c}$. In this
case the expression for dynamical resistivity is described in the form

\begin{align}
\frac{\rho_{xx}}{\rho_0}&\approx\left(\frac{\tau e E_0 v_\text{F}}{2\epsilon_\text{F}} \right)^2J^2_{-1}\left[\frac{eE_0q}{m^\ast\left(\omega^2_c -\omega^2\right)}%
\right]\sum_{n,m,k} J^2_n\left(qk_xl^2\right)  \notag \\
&\times J^2_m\left(2qk_yl^2\right)J^2_k\left[\frac{eE_0q}{%
m^\ast\left(\omega^2_c -\omega^2\right)}\right]\delta_{n+m+k,0},
\label{Eq:ResistivityApprox3}
\end{align}
Note that the static case ($\omega=0$) is dominated by the terms $n=m=k=s=0$
in the sums, whereas all other values of these integers contribute to the
dynamic case. In the regime of intermediate microwave radiation frequency range, $1\ll
\omega\tau\ll\omega_c\tau$ and strong magnetic field, the classical
oscillations take the form
\begin{align}
\frac{\rho_{xx}}{\rho_0}&\approx\left(\frac{e E_0 v_\text{F}}{2\epsilon_\text{F}\omega} \right)^2 J^2_0\left(qk_Fl^2\right)J^4_0\left[\frac{eE_0q}{m^\ast\left(\omega^2_c
-\omega^2\right)}\right].  \label{Eq:ResistivityApprox4}
\end{align}
Fig.~\ref{fig5} demonstrates the microwave radiation frequency-dependence of $\rho
_{xx}/\rho _{0}$ oscillations. It shows that the mechanism of suppression of
SdHO strongly depends on the microwave radiation frequency. We see that high frequency MW
can efficiently drive the system to zero-resistance state as observed in
experiments~\cite{Mani Nat.420-646,ManiAPL-85-4962,HaltePRB121301}.
This is also obvious from Fig.~\ref{fig6} where $\rho_{xx}$ vanishes when the MW frequency is
sufficiently large.
Hence in this regime the system resides in a zero-resistance state.

\section{Summary and Conclusions}\label{Sec:Conc}

In conclusion, we have analyzed a semiclassical theory of magnetotransport in two dimensional
electron gas (2DEG) systems irradiated by microwaves. In this regard, We have
discussed radiation-assisted dynamics of a charged particle in an external magnetic field.
In the presence of a perpendicular magnetic field the resistivity of the system
shows Shubnikov-de Haas oscillations (SdHO). However, if the system is
illuminated by high-frequency microwave radiation, the SdHO are
suppressed and consequently the resistivity of the system vanishes in some
intervals of the magnetic field which are associated with zero-resistance states (ZRS).
Moreover, a detailed investigation of parametric regimes where zero resistance states in
our system can be observed has been performed. Furthermore, experimental relevance with MW
illuminated 2DEG systems has been established.

\section{Acknowledgement}

The authors thank John Schliemann for helpful discussion on the topic. K. Sabeeh would
like to gratefully acknowledge the support of the Abdus Salam International Center for
Theoretical Physics (ICTP), in Trieste, Italy through the Associate Scheme where a part
of the work was completed. Further, support of Higher Education Commission (HEC) of
Pakistan through project No. 20-1484/R \& D/09 is acknowledged.

\end{document}